\documentclass[12pt]{article}

\usepackage{amsfonts}
\usepackage{epsfig}
\usepackage{a4}
\usepackage{theorem}
\usepackage{amssymb}
\usepackage{graphicx}

\newcommand{\be}[1]{\begin{equation}\label{#1}}
\newcommand{\ee}{\end{equation}}
\newcommand{\ba}[1]{\begin{eqnarray}\label{#1}}
\newcommand{\ea}{\end{eqnarray}}

\begin{document}
\title{The pomeron in closed bosonic string theory}
\author{
A.R.~Fazio
\vspace{3mm} \\
\small Departamento de Fisica, Universidad Nacional de Colombia\\
\small Ciudad Universitaria, Bogot\'a, D.C. Colombia\\
\small arfazio@unal.edu.co\vspace{3mm}\\}
\maketitle
\begin{abstract}
We review the features of the pomeron in the S-matrix theory and in quantum field theory. We extend those general properties to the pomeron of closed bosonic string theory in a Minkowskian background. We compute the couplings of the pomeron to the lowest mass levels of closed bosonic string states in flat space. We recognize the deviation from the linearity of the Regge trajectories in a five dimensional anti De Sitter background. 
\end{abstract}
\section{Introduction}
Consider the $2\rightarrow 2$ scattering amplitude for the process 
$$A+B\rightarrow A^\prime +B^\prime.$$
The Lorentz covariance of the S-matrix requires the scattering amplitude to be written in terms of the Mandelstam variables
\begin{equation}
s=(p_A+p_B)^2\,\,\,\,\,\,\, t=(p_A-p_{A^\prime})^2\,\,\,\,\, u=(p_A-p_{B^\prime})^2
\end{equation}
which, for particles of equal masses, are related each others by
\begin{equation}
s+u+t=4m^2.
\end{equation}
The Lorentz invariant scattering amplitude can be written as function of just $s$ and $t$, $\textit{A}(s,t)$, to be studied in the Regge kinematical regime
\begin{equation}
s\approx -u >> m^2\approx-t \approx |\vec{q}|^2\approx E^2\theta^2
\label{reggelimit}
\end{equation}
being $q$ the transferred momentum in the center of mass frame, $p_A-p_{A^\prime}$. $E$ is the center of mass energy of the approaching particles and $\theta$ is the scattering angle. To work in the Regge limit Sudakov's paramenters $s_1,\,\,s_2$ are useful. The transferred momentum gets decomposed as
\begin{equation}
q=\vec{q}-\frac{s_1}{s}p_B+\frac{s_2}{s}p_A
\end{equation}
with the orthogonality relation
\begin{equation}
\vec{q}\cdot p_A = 0 = \vec{q}\cdot p_B.
\label{ortho}
\end{equation}
Putting all masses to zero, as in the high energy regime, we have 
the decomposition
\begin{equation}
q=\vec{q} +\frac{\vec{q}\,\,^2+m^2}{s}(p_A-p_B)
\label{decomposition}
\end{equation}
and, in the Regge limit, we can make the replacement $q\approx \vec{q}$. In the next $\vec{q}$ will be treated as a two dimensional vector, since  $(\ref{ortho})$ implies the orthogonality to the scattering direction. Due to the Sudakov's parametrization we may write
\begin{equation}
d^4 q = \frac{ds_1 ds_2 d^2 q}{2 |s|}.
\end{equation} 
Unitarity of $S$ matrix implies the optical theorem, relating the total cross section to the imaginary part of the forward scattering amplitude. The total cross section is defined as
\begin{equation}
\sigma_\alpha = \int d\beta \frac{d\sigma (\alpha \rightarrow \beta)}{d\beta},
\end{equation}
where $d\beta$ is an infinitesimal phase space element and the optical theorem states that
\begin{equation}
\sigma_\alpha =\frac{\Im A_\alpha(s,0)}{s}.
\label{optical}
\end{equation}
At high energy in the Regge regime it is particularly useful to use the eikonal approximation of the scattering amplitude
\begin{equation}
A(s,t)\approx-2is \int d^2\rho\,\,e^{i\vec{\rho}\cdot \vec{q}}\,\,(e^{i\chi (s,\vec{\rho})}-1),
\end{equation}
being $\chi (s,\vec{\rho})$ the eikonal phase and $\vec{\rho}$ the impact parameter, the two-dimensional vector dual Fourier variable of the transfered momentum $\vec{q}$. From the the optical theorem $(\ref{optical})$ we deduce
\begin{equation}
\sigma_\alpha = 2\int d^2\rho\,\,\Re (1-e^{i\chi (s,\vec{\rho})}),
\end{equation}
coinciding, since we are looking at a $2\rightarrow 2$ process,  with the elastic total cross section
\begin{equation}
\sigma_\alpha ^{el}=\int d^2\rho |1-e^{i\chi (s,\vec{\rho})}|^2.
\end{equation}
To introduce the Regge poles and therefore the pomeron we adopt the the Gribov's approach of representing the scattering amplitude by its Mellin transform
\begin{equation} 
A_p(s,t)= \int^{a+i \infty}_{a-i\infty}\frac{d \omega}{2\pi i}\xi_p(\omega)f_p(\omega,t)s^\omega
\label{Mellin}
\end{equation}
with $a>0$. The integral is performed on a contour made by the straight vertical line at $\Re \omega =a$ and leaving on its right all singularities of the integrand in the $\omega$ complex plane. The contour closes at infinity. Here $A_p(s,t)$ brings an additional quantum number, the signature $p=\pm 1$, explicitly appearing in the signature factor as
\begin{equation}
\xi_p(\omega)=-\frac{p+e^{-i\pi\omega}}{sin \pi\omega}
\label{signature}
\end{equation}
and meaning the parity of the partial waves under the $s\leftrightarrow u$ exchange.
The sinus factor in the denominator of $(\ref{signature})$ suggests that $(\ref{Mellin})$ is the analytical continuation to the complex $\omega$ plane of the partial wave expansion in the $t$-channel for the amplitude
\begin{equation}
A(s,t)=\sum_{l=0}^{+\infty} (2l+1)a_l(t)P_l(cos\theta).
\label{partial}
\end{equation}
Making the Regge limit, $A(s,t)$ looks as in $(\ref{Mellin})$ because $P_l(cos\theta)=P_l\left(1+\frac{2s}{t}\right)$ goes to
$\left(\frac{s}{t}\right)^l$,\,\, and $f_p(\omega,t)$ are the analitical continuation of the partial waves with definite signature. Assuming the partial wave $f_p(\omega,t)$ to have isolated poles, the moving poles, with the usual factorized residue: 
\begin{equation}
f_p(\omega,t)=\frac{\gamma^{p}_{A}(t)\gamma^{p}_{B}(t)}{\omega-\omega_p(t)}
\label{Regge partial}
\end{equation}
and replacing the previous into $(\ref{Mellin})$, we get in the Regge limit
\begin{equation} 
A_p(s,t)\approx \frac{\xi_p(\omega_p(t))\gamma^{p}_{A}(t)\gamma^{p}_{B}(t) s^{\omega_p(t)}}{\Gamma(\omega_p(t))}.
\label{regge amplitude}
\end{equation}
The result $(\ref{regge amplitude})$ is understood as the exchange in the $t$-channel of a spin continous object, the Reggeon, whose spin $\omega_p(t)$ is called Regge trajectory. In linear approximation it depends on just two parameters, the intercept $\omega_p(0)$ and the Regge slope $\alpha^\prime_p$
\begin{equation}
\omega_p(t)= \omega_p(0) + \alpha^\prime_p t.
\end{equation}
For $t>0$ we have particles exchanged in the $t$-channel  when $t=m_i^2$ and $\omega_p(m_i^2)=J_i$, the physical spin. The Reggeon having the vacuum quantum numbers (spin zero, C=+1, P=+1) and positive signature is the Pomeron. The contribution of such a trajectory to the total cross section  is proportional to 
\begin{equation}
s^{\omega(0)-1}\gamma(0),
\end{equation} 
which agrees with the Pomeranchuk theorem of the increasing of the total cross section with energy due to the exchange of the vacuum quantum numbers, if $(\omega(0)-1)>1$. The asymptotic behaviour of scattering amplitudes in the Born approximation is governed by the spin $\sigma$ of the exchanged particle in the t-channel. In quantum chromodynamics such a leading behaviour is given by the so called reggezeid gluon. We just spend few words about the meaning of reggezeid particle \cite{mandel}. When scattering amplitudes calculated by Lagrangian field theory are expressed as $(\ref{Mellin})$ it usually happens that the representation must be modified by the addition of polynomial in the cosine of the scattering angle. A polynomial $Q_n(\cos \theta)$ will only affect the nth partial wave and will therefore leave the function $a_l(t)$ unchanged. The existence of such a term will mean however that the function $a_l(t)$ is not equal to the physical partial wave at $l=n$. In other words the Regge function $f_p(\omega,t)$ will contain Kronecher delta singularities at $l=n$ if it is to represent the physical partial waves correctly. If a theory has an elementary particle of spin $\sigma$, a Kronecher delta singularity will occurr at $l=\sigma$ in the channel of the quantum numbers of the elementary particle. In quantum electrodynamics the photon remains an elementary particle but the electron does Reggeize, the Kronecher delta singularity arising in the Born approximation on account of one-electron exchange in the $t$-channel disappears as a consequence of the radiative corrections. In quantum chromodynamics the gluon and the quarks Reggeize. Especially important is the gluon Reggeization, because of the non-decreasing with energy of the total cross sections are provided by gluon exchanges. At each order of perturbation theory amplitudes with negative signatures do dominate, owing to cancellation of the leading logarithmic terms in the amplitudes with positive signature \cite{lipatov}, \cite{fadin}. The Pomeron emerges as compound of two reggeized gluons. The high energy scattering amplitude for the colourless particle described by the Feynman diagrams containing only two intermediate gluons with momenta $k$ and $q-k$ in the $t$ channel has the Mellin transform
\begin{equation}
f(\omega,q^2) \delta^2(q-q^\prime)=\int \prod \frac{d^2\rho_r d^2\rho_r^\prime}{(2\pi)^4}\Phi(\vec{\rho_1},\vec{\rho_2},\vec{q^\prime})f_\omega(\vec{\rho_1},\vec{\rho_2},\vec{\rho^\prime _1},\vec{\rho^\prime _2})\Phi(\vec{\rho^\prime_1},\vec{\rho^\prime_2},\vec{q^\prime}).
\end{equation}
The impact factors $\Phi$ describe the inner structure of the colliding particles, the quantity $f_\omega(\vec{\rho_1},\vec{\rho_2},\vec{\rho^\prime _1},\vec{\rho^\prime _2})$ can be considered as a four-point gluon Green function in a two-dimensional impact parameter space. It depends on $\rho_{ik}=\rho_i-\rho_k$. Due to the colourless property of the colliding particles the $f_\omega(\vec{\rho_1},\vec{\rho_2},\vec{\rho^\prime _1},\vec{\rho^\prime _2})$ can be recast into Moebius invariant form depending only on the two anharmonic ratios of the vectors $\vec{\rho_1}$, $\vec{\rho_2}$,$\vec{\rho^\prime_1}$,$\vec{\rho^\prime_2}$ chosen as follows
\begin{equation}
\alpha = \left|\frac{\rho_{11^\prime}\rho_{22^\prime}}{\rho_{12^\prime}\rho_{1^\prime 2}}\right|\,\,\,\,\,\,\,\beta = \left|\frac{\rho_{11^\prime}\rho_{22^\prime}}{\rho_{12}\rho_{1^\prime 2^\prime}}\right|
\end{equation}
here $\rho_k = x_k+i y_k \,\,\,{\rho_k}^\ast = x_k - i y_k$ for all two dimensional vectors $\vec{\rho_k}=(x_k,y_k)$. A world-sheet structure is emerging even in the perturbative description of the pomeron. In this paper we are going to explore how in closed bosonic string theory a pomeron description is emerging from pure world-sheet calculations. In section two we review the vertex operator of the pomeron in string theory in flat space and we provide the  coupling of the pomeron to the colliding asymptotic string states. In section three we consider the Regge limit of the scattering amplitude of two dilatons going to two dilatons in an $ADS_5$ background and we confirm the deviation from the linearity of the Regge trajectories as found in $(\cite{polchinski})$. Section four is for conclusions and discussions.

\section{Regge behaviour in string theory in flat space}
Consider a bosonic string moving in a Minkowskian flat background of 26 dimensions. According to the usual notations in string theory the metric tensor is $\eta_{\mu\nu}= diag(-1, 1, 1......)$, and we will be using such a metric tensor in the following. The Moebius fixed Virasoro-Shapiro amplitude for the $2\longrightarrow 2$ closed string tachyons is given by
\begin{equation}
A(s,t)=\int d^2 z <V_1(z,\bar{z})V_2(0)V_3(1)V_4(\infty)>
\label{VS}
\end{equation} 
being $V_i(z,\bar{z})=e^{ip_i X}(z,\bar{z})$.
Performing the calculation of the matrix element in the conformal gauge, the worldsheet action is actually the two-dimensional Klein-Gordon action and therefore (\ref{VS}) is calculated by the contractions
\begin{equation}
<T X^\mu (z_1,\bar{z_1})X^\nu (z_2,\bar{z_2})> = -\frac{\alpha^\prime}{2}\eta^{\mu\nu}ln |z_1-z_2|^2,
\end{equation}
one finds
\begin{equation}
A(s,t)=\int d^2 z |z|^{-4-\frac{\alpha^\prime t}{2}}|1-z|^{-4-\frac{\alpha^\prime s}{2}}.
\end{equation}
It is known that $A(s,t)$ is dominated by a saddle point \cite{gross} which, in the Regge limit $(\ref{reggelimit})$, implies that the integral receives its main contribution from the small $z$ region, $ sz\approx1 $.
After the appropriate analytical continuation to the region $-8<\alpha^\prime t<-4 $ and $s$ continued to be positive imaginary we get 
\begin{equation}
A(s,t)\stackrel{Regge}{\longrightarrow}2\pi\frac{\Gamma (-1-\frac{\alpha^\prime t}{2})}{\Gamma (2+\frac{\alpha^\prime t}{4})}\left(-i \frac{\alpha^\prime s}{4}\right)^{2+\frac{\alpha^\prime t}{2}}.
\label{ReggeVS}
\end{equation}
We may reach the same result by performing the operator product expansion (OPE) in the small Regge $z$ region so the usual OPE $z<<1$
$$e^{ip_1\cdot X}(z,\bar{z})e^{ip_2\cdot X}(0,0)\approx |z|^{-4-\frac{\alpha^\prime t}{2}}:\exp[i(p_1+p_2)X(0,0)]:$$
is replaced by the Regge OPE $z<<1$, $sz \approx 1$
\begin{equation}
|z|^{-4-\frac{\alpha^\prime t}{2}}:\exp[i(p_1+p_2)X(0,0)+ip_1 (z\partial X(0,0)+\bar{z}\bar{\partial}X (0,0))]:
\label{reggeOPE}
\end{equation}
By integrating the previous expression $(\ref{reggeOPE})$ on the entire complex plane one gets the Pomeron vertex operator 
\begin{equation}
V_P(0)=2\pi\frac{\Gamma (-1-\frac{\alpha^\prime t}{2})}{\Gamma (2+\frac{\alpha^\prime t}{4})}e^{-i\pi-i\pi\frac{\alpha' t}{4}}:\exp[i(p_1+p_2)X(0,0)](p_1\cdot\partial X p_1\cdot\bar{\partial} X)^{1+\frac{\alpha^\prime t}{4}}:
\label{pomeron}
\end{equation} 
By replacing ($\ref{pomeron}$) into ($\ref{VS}$) one gets the complete Regge limit $(\ref{ReggeVS})$ of the Virasoro-Shapiro amplitude. This procedure shows that the the scattering amplitude factorizes on the pomeron vertex operator, which could be interpreted as a string state of continuous spin $\alpha(t)=2+\frac{\alpha^\prime t}{4}$, saying therefore that it is not a physical particle but a Regge trajectories. The Regge exponent and so the spin of the pomeron is such that this vertex operator satisfies the physical state condition
\begin{equation}
L_0=1.
\end{equation}

We now generalize the four tachyons result to any 2 by 2 scattering amplitude. In the general case the integration at operator level gives
\begin{equation}
\int d^2 z V_1(z,\bar{z}) V_2(0,0)\stackrel{Regge}{\sim}C_{12}V_P(0),
\label{coupling}
\end{equation}
$C_{12}$ are momentum dependent symmetric constants, bringing the polarization tensor terms in the two operators. In the following we will compute the coupling $C_{12}$ for the vertex operators of mass levels $m^2=-\frac{4}{\alpha^\prime},\,0,\,\,\frac{4}{\alpha^\prime}$ and $m^2=\frac{8}{\alpha^\prime}$. The list of such a vertex operators is done in \cite{weinberg}. Here the vertex operators will be listed in pairs, together with their masses, the constraints on their polarization tensors and with their coupling to the pomeron.

\begin{eqnarray}
	V_1(z,\bar{z})&=& :e^{ip_1\cdot X}(z,\bar{z}): \,\,\,\,\,\, V_2(0,0)=:e^{ip_2\cdot X}(0,0):\nonumber\\
	m_1^2&=& m_2^2=-\frac{4}{\alpha^\prime}\nonumber\\
	C_{12} &=& 1
\end{eqnarray}

\begin{eqnarray}
	V_1(z,\bar{z})&=& :e^{ip_1\cdot X}(z,\bar{z}): \,\,\,\,\,\, V_2(0,0)=e_{\mu\nu}(p_2):\partial X^\mu \bar{\partial}X^\nu e^{ip_2\cdot X}(0,0):\nonumber\\
	m_2^2 &=& 0\,\,\,\,\,\,\,{{p_2}^\mu}e_{\mu;\nu}(p_2)={{p_2}^\nu} e_{\mu;\nu}(p_2)=0\nonumber\\
	C_{12} &=& -\frac{{\alpha^\prime}^2 p_1^\mu p_1^\nu e_{\mu\nu}(p_2)}{4}
\end{eqnarray}

\begin{eqnarray}
	V_1(z,\bar{z})&=& :e^{ip_1\cdot X}(z,\bar{z}): \,\,\,\,\,\, V_2(0,0)=e_{\mu\nu;\lambda\rho}(p_2):\partial X^\mu \partial X^\nu\bar{\partial}X^\lambda\bar{\partial}X^\rho e^{ip_2\cdot X}(0,0):\nonumber\\
m_2^2 &=& \frac{4}{\alpha^\prime}\,\,\,\,\,\, {p_2}^\mu e_{\mu\nu;\lambda\rho}(p_2)={p_2}^\lambda e_{\mu\nu;\lambda\rho}(p_2)= 0 
	\,\,\,\,\,\,\eta^{\mu\nu}e_{\mu\nu;\lambda\rho}(p_2)=0=\eta^{\lambda\rho}e_{\mu\nu;\lambda\rho}(p_2)\nonumber\\
	C_{12} &=&\frac{{\alpha^\prime}^4 p_1^\mu p_1^\nu p_1^\rho p_1^\lambda e_{\mu\nu;\lambda\rho}(p_2)}{16}
\end{eqnarray}

\begin{eqnarray}
V_1(z,\bar{z})&=& e^{ip_1\cdot X}(z,\bar{z}) \,\,\,\,\,\, V_2(0,0)=e_{\mu\nu\lambda;\rho\sigma\tau}(p_2):\partial X^\mu \partial X^\nu\partial X^\lambda\bar{\partial}X^\sigma\bar{\partial}X^\rho \bar{\partial}X^\tau e^{ip_2\cdot X}(0,0):\nonumber\\
m_2^2 &=& \frac{8}{\alpha^\prime}\,\,\,\,\,\,\,\,\,\,\,{p_2}^\mu e_{\mu\nu\sigma;\lambda\rho\tau}(p_2)={p_2}^\lambda e_{\mu\nu\sigma;\lambda\rho\tau}(p_2)= 0\,\,\,\,\,\,\,\eta^{\mu\nu}e_{\mu\nu\lambda;\rho\sigma\tau}(p_2)=0=\eta^{\rho\sigma}e_{\mu\nu\lambda;\rho\sigma\tau}(p_2)\nonumber\\
	C_{12} &=&-\frac{{\alpha^\prime}^6 p_1^\mu p_1^\nu p_1^\rho p_1^\lambda p_1^\sigma p_1^\tau e_{\mu\nu\lambda;\rho\sigma\tau}(p_2)}{64}
\end{eqnarray}

\begin{eqnarray}
V_1(z,\bar{z})&=& e^{ip_1\cdot X}(z,\bar{z})\nonumber\\
V_2(0,0)&=& e_{\mu\nu;\rho\sigma\tau}(p_2):[\partial X^\mu \partial^2 X^\nu\bar{\partial}X^\sigma\bar{\partial}X^\rho\bar{\partial}X^\tau \pm \bar{\partial} X^\mu \bar{\partial}^2 X^\nu{\partial}X^\sigma{\partial}X^\rho{\partial}X^\tau] e^{ip_2\cdot X}(0,0):\nonumber\\
m_2^2 &=& \frac{8}{\alpha^\prime}\,\,\,\,\,\,e_{\mu\nu;\rho\sigma\tau}(p_2)=-e_{\nu\mu;\rho\sigma\tau}(p_2)\,\,\,\,\,\,\,\eta^{\rho\sigma}e_{\mu\nu;\rho\sigma\tau}(p_2)=0\,\,\,\,\,\,\,\,{p_2}^\mu e_{\mu\nu;\rho\sigma\tau}(p_2)={{p_2}^\rho} e_{\mu\nu;\rho\sigma\tau}(p_2)=0\nonumber\\
C_{12} &=& 0
\end{eqnarray}

\begin{eqnarray}
	V_1(z,\bar{z})&=&:e^{ip_1\cdot X}(z,\bar{z}): \,\,\,\,\,\, V_2(0,0)=e_{\mu\nu;\rho\sigma}(p_2):[\partial X^\mu \partial^2 X^\nu\bar{\partial}X^\rho\bar{\partial}^2 X^\sigma] e^{ip_2\cdot X}(0,0):\nonumber\\
m_2^2 &=& \frac{8}{\alpha^\prime}\,\,\,\,\,\,e_{\mu\nu;\rho\sigma}(p_2)=-e_{\nu\mu;\rho\sigma}(p_2)=-e_{\nu\mu;\sigma\rho}(p_2)\,\,\,\,\,\,\,{p_2}^\mu e_{\mu\nu;\rho\sigma}(p_2)={{p_2}^\rho} e_{\mu\nu;\rho\sigma}(p_2)=0\nonumber\\	
C_{12} &=&0
\end{eqnarray}

\begin{eqnarray}
	V_1(z,\bar{z})&=& e_{\mu\nu}(p_1):\partial X^\mu \bar{\partial}X^\nu e^{ip_1\cdot X}(z,\bar{z}): \,\,\,\,\,\, V_2(0,0)=:e_{\alpha\beta}(p_2)\partial X^\alpha \bar{\partial}X^\beta e^{ip_2\cdot X}(0,0):\nonumber\\
	C_{12} &=&\frac{{\alpha^\prime}^4 p_2^\mu p_2^\nu e_{\mu\nu}(p_1) p_1^\alpha p_1^\beta e_{\alpha\beta}(p_2)}{16}+\frac{{\alpha^\prime}^3 [p_2^\mu {e_{\mu}}^{\nu}(p_1)p_1^\alpha e_{\alpha\nu}(p_2)+{e^{\mu}}_{\nu}(p_1)p_2^\nu e_{\mu\beta}(p_2)p_1^\beta]}{8}\nonumber\\
&+&\frac{{\alpha^\prime}^2 e_{\mu\nu}(p_1) e^{\mu\nu}(p_2)}{4}
\end{eqnarray}

\begin{eqnarray}
	V_1(z,\bar{z})&=& e_{\mu\nu}(p_1):\partial X^\mu \bar{\partial}X^\nu e^{ip_1\cdot X}(z,\bar{z}): \,\,\,\,\,\,V_2(0,0)=e_{\mu\nu;\lambda\rho}(p_2):\partial X^\mu \partial X^\nu\bar{\partial}X^\lambda\bar{\partial}X^\rho e^{ip_2\cdot X}(0,0):
	 \nonumber\\
	C_{12} &=&-\frac{{\alpha^\prime}^6 p_2^\alpha e_{\alpha\beta}(p_1)p_2^\beta p_1^\mu p_1^\nu e_{\mu\nu;\lambda\rho}(p_2)p_1^\lambda p_1^\rho}{64}-\frac{{\alpha^\prime}^5 p_2^\alpha e_{\alpha\beta}(p_1)p_1^\mu p_1^\nu {{e_{\mu\nu;}}^{\beta}}_{\lambda}(p_1)p_1^\lambda}{16}\nonumber\\
&&-\frac{{\alpha^\prime}^5 p_2^\alpha e_{\alpha\beta}(p_1)p_1^\mu  {{e_{\mu}}^{\beta}}_{;\lambda\rho}(p_2)p_1^\lambda p_1^\rho}{16}-\frac{{\alpha^\prime}^4 p_1^\alpha e_{\alpha\beta;\rho\lambda}(p_2) e^{\beta\rho}(p_1)p_1^\lambda}{4}
\end{eqnarray}

\begin{eqnarray}
	V_1(z,\bar{z})&=& e_{\mu\nu}(p_1):\partial X^\mu \bar{\partial}X^\nu e^{ip_1\cdot X}(z,\bar{z}):\nonumber\\
	V_2(0,0)&=&e_{\alpha\beta\lambda;\rho\sigma\tau}(p_2):\partial X^\alpha \partial X^\beta\partial X^\lambda\bar{\partial}X^\rho\bar{\partial}X^\sigma\bar{\partial}X^\tau e^{ip_2\cdot X}(0,0):\\
	C_{12} &=&\frac{{\alpha^\prime}^8 p_2^\mu e_{\mu\nu}(p_1)p_2^\nu p_1^\alpha p_1^\beta p_1^\lambda e_{\alpha\beta\lambda;\rho\sigma\tau}(p_2)p_1^\rho p_1^\sigma p_1^\tau}{256}-\frac{3{\alpha^\prime}^7 p_2^\mu e_{\mu\nu}(p_1){e^{\nu}}_{\beta\lambda;\rho\sigma\tau}(p_2)p_1^\beta p_1^\lambda p_1^\rho p_1^\sigma p_1^\tau}{128}\nonumber\\
	&&-\frac{3{\alpha^\prime}^7 p_2^\mu e_{\mu\nu}(p_1){{e_{\alpha\beta\lambda;}}^{\nu}}_{\sigma\tau}(p_2)p_1^\alpha p_1^\beta p_1^\lambda p_1^\sigma p_1^\tau}{128}+\frac{9{\alpha^\prime}^6 p_1^\alpha p_1^\beta {{{{e_{\alpha\beta}}^{\lambda}}_{;}}^{\rho}}_{\sigma\tau}(p_2)p_1^\sigma p_1^\tau e_{\lambda\rho}(p_1)}{64}\nonumber
\end{eqnarray}

\begin{eqnarray}
V_1(z,\bar{z})&=& e_{\mu\nu}(p_1):\partial X^\mu \bar{\partial}X^\nu e^{ip_1\cdot X}(z,\bar{z}):\nonumber\\
V_2(0,0)&=& e_{\mu\nu;\rho\sigma\tau}(p_2):[\partial X^\mu \partial^2 X^\nu\bar{\partial}X^\sigma\bar{\partial}X^\rho\bar{\partial}X^\tau \pm \bar{\partial} X^\mu \bar{\partial}^2 X^\nu{\partial}X^\sigma{\partial}X^\rho{\partial}X^\tau] e^{ip_2\cdot X}(0,0):\nonumber\\
C_{12} &=& -\frac{3 i {\alpha^\prime}^6}{32}[p_1^\mu e_{\mu\nu;\rho\sigma\tau}(p_2)p_1^\rho p_1^\sigma p_1^\tau {e^{\nu}}_\beta(p_1) p_2 ^\beta \pm  p_1^\mu e_{\mu\nu;\rho\sigma\tau}(p_2)p_1^\rho p_1^\sigma p_1^\tau {e_{\alpha}}^\nu(p_1) p_2 ^\alpha] \nonumber\\
&&+\frac{9i{\alpha^\prime}^5}{8}[p_1^\mu e_{\mu\nu;\rho\sigma\tau}(p_2)p_1^\rho p_1^\sigma {e^{\nu\tau}}(p_1)\pm  p_1^\mu e_{\mu\nu;\rho\sigma\tau}(p_2)p_1^\rho p_1^\sigma e^{\tau\nu}(p_1)]
\end{eqnarray}

\begin{eqnarray}
V_1(z,\bar{z})&=& e_{\mu\nu}(p_1):\partial X^\mu \bar{\partial}X^\nu e^{ip_1\cdot X}(z,\bar{z}): \,\,\,\,\,\,V_2(0,0)=e_{\mu\nu;\rho\sigma}(p_2):[\partial X^\mu \partial^2 X^\nu\bar{\partial}X^\rho\bar{\partial}^2 X^\sigma] e^{ip_2\cdot X}(0,0): \nonumber \\
C_{12} &=&-\frac{{\alpha^\prime}^4 e^{\mu\rho}(p_1)e_{\mu\nu;\rho\sigma}(p_2)p_1^\sigma p_1^\nu}{16}
\end{eqnarray}

\begin{eqnarray}
V_1(z,\bar{z})&=& e_{\mu\nu;\lambda\rho}(p_1):\partial X^\mu \partial X^\nu \bar{\partial}X^\lambda \bar{\partial}X^\rho e^{ip_1\cdot X}(z,\bar{z}):
	\nonumber\\
V_2(0,0)&=& e_{\alpha\beta;\gamma\delta}(p_2):\partial X^\alpha \partial X^\beta\bar{\partial}X^\gamma\bar{\partial}X^\delta e^{ip_2\cdot X}(0,0):
	 \nonumber\\
C_{12}&=& \frac{{\alpha^\prime}^8 p_1^\alpha p_1^\beta e_{\alpha\beta;\gamma\delta}(p_2)p_1^\gamma p_1^\delta p_2^\mu p_2^\nu e_{\mu\nu;\lambda\rho}(p_1)p_2^\lambda p_2^\rho}{256}-\frac{{\alpha^\prime}^7 p_1^\alpha p_1^\beta e_{\alpha\beta;\gamma\delta}(p_2)p_1^\gamma p_2^\mu p_2^\nu {e_{\mu\nu;\lambda}}^{\delta}(p_1)p_2^\lambda}{32}\nonumber\\
&&-\frac {{\alpha^\prime}^7 p_1^\beta e_{\alpha\beta;\gamma\delta}(p_2)p_1^\gamma p_1^\delta p_2^\nu {e^{\alpha}}_{\nu;\lambda\rho}(p_1)p_2^\lambda p_2^\rho}{32}+ \frac{{\alpha^\prime}^6 p_2^\mu p_2^\nu e_{\mu\nu;\lambda\gamma}(p_1)p_1^\alpha p_1^\beta {e_{\alpha\beta;}}^{\lambda\gamma}(p_2)}{32}\nonumber\\
&&+ \frac{{\alpha^\prime}^6 e_{\mu\nu;\lambda\rho}(p_1)p_2^\lambda p_2^\rho  {e^{\mu\nu}}_{;\gamma\delta}(p_2)p_1^\gamma p_1^\delta}{32}  +\frac{{\alpha^\prime}^6 p_2^\mu e_{\mu\nu;\lambda\rho}(p_1)p_2^\lambda p_1^\alpha  
{{{{{e_{\alpha}}^{\nu}}_{;}}^{\rho}}_{\gamma}}(p_2)p_1^\gamma}{16}\\
&&-\frac{{\alpha^\prime}^5 p_1^\alpha e_{\alpha\beta;\gamma\delta}(p_2)p_2^\mu {{{e_{\mu}}^{\beta}}_{;}}^{\gamma\delta}(p_1)}{4}
-\frac{{\alpha^\prime}^5 p_2^\lambda e_{\mu\nu;\lambda\rho}(p_1)p_1^\gamma {{{{e^{\mu\nu}}}}_{;\gamma}}^{\rho}(p_2)}{4}+ \frac{{\alpha^\prime}^4 e_{\mu\nu;\lambda\rho}(p_1)e^{\mu\nu;\lambda\rho}(p_2)}{4}\nonumber
\end{eqnarray}

\begin{eqnarray}
V_1(z,\bar{z})&=& e_{\mu\nu;\lambda\rho}(p_1):\partial X^\mu \partial X^\nu \bar{\partial}X^\lambda \bar{\partial}X^\rho e^{ip_1\cdot X}(z,\bar{z}):\nonumber\\
V_2(0,0)&=& e_{\mu\nu\psi;\varphi\eta\chi}(p_2):\partial X^\mu \partial X^\nu\partial X^\psi\bar{\partial}X^\varphi\bar{\partial}X^\eta\bar{\partial}X^\chi e^{ip_2\cdot X}(0,0):
	 \nonumber\\
C_{12}&=& -\frac{{\alpha^\prime}^{10} p_2^\alpha p_2^\beta e_{\alpha\beta;\gamma\delta}(p_1) p_2^\gamma p_2^\delta p_1^\mu p_1^\nu p_1^\psi e_{\mu\nu\psi;\varphi\eta\chi}(p_2)p_1^\chi p_1^\varphi p_1^\eta}{1024}\\
&&+\frac{3{\alpha^\prime}^{9} p_2^\alpha p_2^\beta e_{\alpha\beta;\gamma\delta}(p_1) p_2^\gamma p_1^\mu p_1^\nu p_1^\psi {{e_{\mu\nu\psi;\varphi}}^\delta}_\chi(p_2)p_1^\chi p_1^\varphi}{256}\nonumber\\
&&
+ \frac{3{\alpha^\prime}^{9}p_2^\beta e_{\alpha\beta;\gamma\delta}(p_1) p_2^\gamma p_2^\delta p_1^\mu p_1^\psi {{e_{\mu}}^{\alpha}}_{\psi;\varphi\eta\chi}(p_2)p_1^\chi p_1^\varphi p_1^\eta}{256}-\frac{3{\alpha^\prime}^{8} p_2^\alpha p_2^\beta e_{\alpha\beta;\gamma\delta}(p_1) p_1^\mu p_1^\nu p_1^\psi {{e_{\mu\nu\psi;}}^{\gamma\delta}}_{\chi}(p_2)p_1^\chi}{128}\nonumber\\
&&-\frac{9{\alpha^\prime}^{8} p_2^\alpha e_{\alpha\beta;\gamma\delta}(p_1)p_2^\gamma p_1^\mu p_1^\psi {{{{e_{\mu}}^{\beta}}_{\psi;\varphi}}^{\delta}}_{\chi}(p_2)p_1^\chi p_1^\varphi}{128}-\frac{3{\alpha^\prime}^{8} e_{\alpha\beta;\gamma\delta}(p_1)p_2^\gamma p_2^\delta p_1^\psi {e^{\alpha\beta}}_{\psi;\varphi\eta\chi}(p_2)p_1^\chi p_1^\varphi p_1^\eta}{128}\nonumber\\
&&+\frac{9{\alpha^\prime}^{7} p_2^\alpha e_{\alpha\beta;\gamma\delta}(p_1)p_1^\mu p_1^\psi {{{{e_{\mu}}^{\beta}}_{\psi;}}^{\gamma\delta}}_{\chi}(p_2)p_1^\chi}{32}+\frac{9{\alpha^\prime}^{7} e_{\alpha\beta;\gamma\delta}(p_1)p_2^\gamma p_1^\psi {{{e^{\alpha\beta}}_{\psi;\varphi}}^{\delta}}_{\chi}(p_2)p_1^\varphi p_1^\chi}{32}\nonumber\\
&&-\frac{9{\alpha^\prime}^{6} e_{\alpha\beta;\gamma\delta}(p_1)p_1^\psi {{{e^{\alpha\beta}}_{\psi;}}^{\gamma\delta}}_{\chi}(p_2)p_1^\chi}{16}\nonumber
\end{eqnarray}

\begin{eqnarray}
V_1(z,\bar{z})&=& e_{\mu\nu;\lambda\rho}(p_1):\partial X^\mu \partial X^\nu \bar{\partial}X^\lambda \bar{\partial}X^\rho e^{ip_1\cdot X}(z,\bar{z}):\nonumber\\
V_2(0,0)&=& e_{\mu\nu;\rho\sigma}(p_2):[\partial X^\mu \partial^2 X^\nu\bar{\partial}X^\rho\bar{\partial}^2 X^\sigma] e^{ip_2\cdot X}(0,0):\nonumber \\
C_{12} &=& \frac{{\alpha^\prime}^{6}p_1^\alpha e_{\alpha\beta;\gamma\delta}(p_2)p_1^\gamma p_2^\mu {{{e_{\mu}}^{\beta}}_{;\lambda}}^{\delta}(p_1)p_2^\lambda}{8}
\end{eqnarray}

\begin{eqnarray}
V_1(z,\bar{z})&=& e_{\mu\nu;\lambda\rho}(p_1):\partial X^\mu \partial X^\nu \bar{\partial}X^\lambda \bar{\partial}X^\rho e^{ip_1\cdot X}(z,\bar{z}):\nonumber\\ 
V_2(0,0)&=&e_{\mu\nu;\rho\sigma\tau}(p_2):[\partial X^\mu \partial^2 X^\nu\bar{\partial}X^\sigma\bar{\partial}X^\rho\bar{\partial}X^\tau \pm \bar{\partial} X^\mu \bar{\partial}^2 X^\nu{\partial}X^\sigma{\partial}X^\rho{\partial}X^\tau] e^{ip_2\cdot X}(0,0):\nonumber\\
C_{12} &=& e_{\mu\nu;\lambda\rho}(p_1)e_{\alpha\beta;\gamma\delta\epsilon}(p_2)\left(-\frac{i {\alpha^\prime}^{8}p_2^{\lambda}p_2^{\rho}p_2^{\nu}p_1^{\delta}p_1^{\beta}p_1^{\gamma}p_1^{\epsilon}\eta^{\mu\alpha}}{128}\mp \frac{i {\alpha^\prime}^{8}p_2^{\mu}p_2^{\rho}p_2^{\nu}p_1^{\delta}p_1^{\beta}p_1^{\gamma}p_1^{\epsilon}\eta^{\lambda\alpha}}{128}\right.\nonumber\\
&&\left.+\frac{3i {\alpha^\prime}^{7}p_2^{\mu}p_2^{\lambda}p_1^{\delta}p_1^{\alpha}p_1^{\gamma}\eta^{\nu\beta}\eta^{\rho\epsilon}}{64}\pm \frac{3i {\alpha^\prime}^{7}p_2^{\mu}p_2^{\lambda}p_1^{\delta}p_1^{\alpha}\eta^{\rho\beta}\eta^{\nu\epsilon}}{64}\right)
\end{eqnarray}

\begin{eqnarray}
V_1(z,\bar{z})&=& e_{\mu\nu\psi;\varphi\eta\chi}(p_1):\partial X^\mu \partial X^\nu \partial X^\psi \bar{\partial}X^\varphi \bar{\partial}X^\upsilon\bar{\partial}X^\chi e^{ip_1\cdot X}(z,\bar{z}):\nonumber\\
V_2(0,0)&=& e_{\alpha\beta\lambda;\rho\sigma\tau}(p_2):[\partial X^\alpha \partial X^\beta \partial X^\lambda\bar{\partial}X^\rho\bar{\partial}X^\sigma\bar{\partial}X^\tau] e^{ip_2\cdot X}(0,0):\nonumber\\
C_{12}&=& e_{\mu\nu\psi;\varphi\upsilon\chi}(p_1)e_{\alpha\beta\lambda;\rho\sigma\tau}(p_2) \left[\frac{{\alpha^\prime}^{12}p_1^{\alpha}p_1^{\beta}p_1^{\lambda}p_1^{\rho}p_1^{\sigma}p_1^{\tau}p_2^{\mu}p_2^{\nu}p_2^{\psi}p_2^{\chi}p_2^{\varphi}p_2^{\upsilon}}{4096}\right.\nonumber\\
&& -\frac{9 {\alpha^\prime}^{11}}{2048}(p_1^{\alpha}p_1^{\beta}p_1^{\lambda}p_1^{\rho}p_2^{\mu}p_2^{\nu}p_2^{\psi}p_2^{\chi}p_2^{\varphi}\eta^{\sigma\upsilon}+ p_1^{\alpha}p_1^{\lambda}p_1^{\rho}p_1^{\sigma}p_2^{\mu}p_2^{\psi}p_2^{\upsilon}p_2^{\varphi}p_2^{\chi}\eta^{\nu\beta})\nonumber\\
&& + \frac{9 {\alpha^\prime}^{10}}{512}\left(p_1^{\alpha}p_1^{\beta}p_1^{\lambda}p_1^{\tau}p_2^{\mu}p_2^{\nu}p_2^{\psi}p_2^{\chi}\eta^{\rho\varphi}\eta^{\sigma\upsilon}+ \frac{9}{2}p_1^{\alpha}p_1^{\lambda}p_1^{\rho}p_1^{\tau}p_2^{\mu}p_2^{\psi}p_2^{\chi}p_2^{\varphi}\eta^{\nu\beta}\eta^{\sigma\upsilon}+ p_1^{\rho}p_1^{\sigma}p_1^{\lambda}p_1^{\tau}p_2^{\varphi}p_2^{\upsilon}p_2^{\psi}p_2^{\chi}\eta^{\mu\alpha}\eta^{\nu\beta}\right)\nonumber\\
&& - \frac{3 {\alpha^\prime}^{9}}{256}(p_1^{\alpha}p_1^{\beta}p_1^{\lambda}p_1^{\mu}p_2^{\nu}p_2^{\psi}\eta^{\chi\tau}\eta^{\rho\varphi}\eta^{\sigma\upsilon}+ p_1^{\sigma}p_1^{\rho}p_1^{\tau}p_2^{\chi}p_2^{\varphi}p_2^{\upsilon}\eta^{\alpha\mu}\eta^{\nu\beta}\eta^{\lambda\psi}+ 27 p_1^{\alpha}p_1^{\tau}p_1^{\lambda}p_2^{\psi}p_2^{\mu}p_2^{\chi}\eta^{\sigma\upsilon}\eta^{\nu\beta}\eta^{\varphi\rho}\nonumber\\&&+ 27 p_1^{\rho}p_1^{\tau}p_1^{\lambda}p_2^{\psi}p_2^{\varphi}p_2^{\chi}\eta^{\sigma\upsilon}\eta^{\nu\beta}\eta^{\alpha\mu})\nonumber\\&&+ \frac{27 {\alpha^\prime}^{8}}{64}\left(\frac{1}{2}p_1^{\alpha}p_1^{\lambda}p_2^{\mu}p_2^{\psi}\eta^{\nu\beta}\eta^{\chi\tau}\eta^{\rho\varphi}\eta^{\sigma\upsilon}+ 3 p_1^{\lambda}p_1^{\tau}p_2^{\psi}p_2^{\chi}\eta^{\nu\beta}\eta^{\mu\alpha}\eta^{\rho\varphi}\eta^{\sigma\upsilon}+\frac{1}{2} p_1^{\rho}p_1^{\tau}p_2^{\chi}p_2^{\varphi}\eta^{\nu\beta}\eta^{\mu\alpha}\eta^{\lambda\psi}\eta^{\sigma\upsilon}\right)\nonumber\\
&&\left.-\frac{27 {\alpha^\prime}^{7}}{32}(p_1^{\lambda}p_2^{\psi}\eta^{\nu\beta}\eta^{\chi\tau}\eta^{\rho\varphi}\eta^{\sigma\upsilon}\eta^{\mu\alpha}+p_1^{\tau}p_2^{\chi}\eta^{\nu\beta}\eta^{\lambda\psi}\eta^{\rho\varphi}\eta^{\sigma\upsilon}\eta^{\mu\alpha})+
\frac{9 {\alpha^\prime}^{6}}{16}\eta^{\nu\beta}\eta^{\lambda\psi}\eta^{\rho\varphi}\eta^{\sigma\upsilon}\eta^{\mu\alpha}\eta^{\chi\tau}\right]
\end{eqnarray}

\begin{eqnarray}
V_1(z,\bar{z})&=& e_{\mu\nu\psi;\varphi\eta\chi}(p_1):\partial X^\mu \partial X^\nu \partial X^\psi \bar{\partial}X^\varphi \bar{\partial}X^\upsilon\bar{\partial}X^\chi e^{ip_1\cdot X}(z,\bar{z}):\nonumber\\
V_2(0,0)&=& e_{\mu\nu;\rho\sigma\tau}(p_2):[\partial X^\mu \partial^2 X^\nu\bar{\partial}X^\sigma\bar{\partial}X^\rho\bar{\partial}X^\tau \pm \bar{\partial} X^\mu \bar{\partial}^2 X^\nu{\partial}X^\sigma{\partial}X^\rho{\partial}X^\tau] e^{ip_2\cdot X}(0,0):\nonumber\\
C_{12} &=& e_{\mu\nu\psi;\varphi\upsilon\chi}(p_1)e_{\alpha\beta;\gamma\delta\epsilon}(p_2)\left[\frac{3i {\alpha^\prime}^{10}}{1024}(p_1^\delta p_1^\epsilon p_1^\gamma p_1^\alpha p_2^\chi p_2^\varphi p_2^\upsilon p_2^\psi p_2^\mu \eta^{\nu\beta} \pm p_1^\epsilon p_1^\gamma p_1^\alpha p_1^{\delta} p_2^\chi p_2^\varphi p_2^\psi p_2^\mu p_2^\nu \eta^{\upsilon\beta})\right.\nonumber\\
&&\left.-\frac{81 i {\alpha^\prime}^{9}}{512}(p_2^\chi p_2^\psi p_2^\varphi p_2^\mu p_1^\delta p_1^\gamma p_1^\alpha \eta^{\nu\beta}\eta^{\gamma\upsilon}\pm p_2^\chi p_2^\psi p_2^\varphi p_2^\mu p_1^\delta p_1^\gamma p_1^\alpha \eta^{\nu\epsilon}\eta^{\beta\upsilon})\right]
\end{eqnarray}

\begin{eqnarray}
V_1(z,\bar{z})&=& e_{\mu\nu\psi;\varphi\eta\chi}(p_1):\partial X^\mu \partial X^\nu \partial X^\psi \bar{\partial}X^\varphi \bar{\partial}X^\upsilon\bar{\partial}X^\chi e^{ip_1\cdot X}(z,\bar{z}):\nonumber\\
V_2(0,0)&=& e_{\mu\nu;\rho\sigma}(p_2):[\partial X^\mu \partial^2 X^\nu\bar{\partial}X^\rho\bar{\partial}^2 X^\sigma] e^{ip_2\cdot X}(0,0):\\
C_{12}&=& -\frac{81}{256}{\alpha^\prime}^{8}p_1^\mu e_{\mu\nu;\gamma\delta}(p_2)p_1^\gamma p_2^\alpha {{{{{{e_{\alpha}}^\nu}_{\lambda;\rho}}^\delta}_\tau}(p_1)p_2^\tau p_2^\rho}+\frac{27}{32}{\alpha^\prime}^{7}p_1^\mu {{e_{\mu}}^\beta}_{;\gamma\delta}(p_2) p_2^\alpha p_2^\lambda {{e_{\alpha\beta\lambda;}}^{\gamma\delta}}_\tau(p_1)p_2^\tau \nonumber
\end{eqnarray}

\begin{eqnarray}
&& V_1(z,\bar{z})= e_{\mu\nu;\rho\sigma\tau}(p_1):[\partial X^\mu \partial^2 X^\nu\bar{\partial}X^\sigma\bar{\partial}X^\rho\bar{\partial}X^\tau \pm \bar{\partial} X^\mu \bar{\partial}^2 X^\nu{\partial}X^\sigma{\partial}X^\rho{\partial}X^\tau] e^{ip_1\cdot X}(z,\bar{z}):\nonumber\\
&& V_2(0,0)= e_{\alpha\beta;\gamma\delta\epsilon}(p_2):[\partial X^\mu \partial^2 X^\nu\bar{\partial}X^\sigma\bar{\partial}X^\rho\bar{\partial}X^\tau \pm \bar{\partial} X^\mu \bar{\partial}^2 X^\nu{\partial}X^\sigma{\partial}X^\rho{\partial}X^\tau] e^{ip_1\cdot X}(0,0):\\
&& C_{12}= e_{\mu\nu;\rho\sigma\tau}(p_1)e_{\alpha\beta;\gamma\delta\epsilon}(p_2)\left[\frac{7{\alpha^\prime}^{9}}{256}p_2^\sigma p_2^\tau p_2^\rho p_2^\mu p_1^\delta p_1^\epsilon p_1^\gamma p_1^\alpha \eta^{\nu\beta}+\left(-\frac{63}{256}\mp\frac{9}{256}\right){\alpha^\prime}^8 p_2^\sigma p_2^\rho p_2^\mu p_1^\delta p_1^\gamma p_1^\alpha \eta^{\nu\beta}\eta^{\tau\epsilon}\right]\nonumber
\end{eqnarray}

\begin{eqnarray}
V_1(z,\bar{z})&=& e_{\mu\nu;\rho\sigma\tau}(p_1):[\partial X^\mu \partial^2 X^\nu\bar{\partial}X^\sigma\bar{\partial}X^\rho\bar{\partial}X^\tau \pm \bar{\partial} X^\mu \bar{\partial}^2 X^\nu{\partial}X^\sigma{\partial}X^\rho{\partial}X^\tau] e^{ip_1\cdot X}(z,\bar{z}):\nonumber\\
V_2(0,0)&=& e_{\mu\nu;\rho\sigma}(p_2):[\partial X^\mu \partial^2 X^\nu\bar{\partial}X^\rho\bar{\partial}^2 X^\sigma] e^{ip_2\cdot X}(0,0):\nonumber\\
C_{12}&=& \frac{i {\alpha^\prime}^{7}}{128}e_{\mu\nu;\rho\sigma\tau}(p_1)e_{\alpha\beta;\gamma\delta}(p_2)(p_2^\sigma p_2^\rho p_2^\mu p_1^\alpha p_1^\gamma \eta^{\nu\beta}\eta^{\tau\delta}\pm p_2^\sigma p_2^\rho p_2^\mu p_1^\alpha p_1^\gamma \eta^{\nu\delta}\eta^{\tau\beta})
\end{eqnarray}

\begin{eqnarray}
V_1(z,\bar{z})&=& e_{\mu\nu;\rho\sigma}(p_1):[\partial X^\mu \partial^2 X^\nu\bar{\partial}X^\rho\bar{\partial}^2 X^\sigma] e^{ip_1\cdot X}(z,\bar{z}):\nonumber\\
V_2(0,0)&=& e_{\mu\nu;\rho\sigma}(p_2):[\partial X^\mu \partial^2 X^\nu\bar{\partial}X^\rho\bar{\partial}^2 X^\sigma] e^{ip_2\cdot X}(0,0):\nonumber\\
C_{12}&=& \frac{25}{32}{\alpha^\prime}^{6}p_2^\mu e_{\mu\nu;\rho\tau}(p_1)p_2^\rho p_1^\alpha {{{e_{\alpha}}^\nu}_{;\gamma}}^\tau (p_2)p_1^\gamma -\frac{35}{16}{\alpha^\prime}^{5}p_2^\mu e_{\mu\nu;\rho\tau}(p_1)p_1^\alpha {e_{\alpha}}^{\nu;\rho\tau}(p_2)\nonumber\\
&&-\frac{35}{16}{\alpha^\prime}^{5}p_2^\rho e_{\mu\nu;\rho\tau}(p_1){{e^{\mu\nu}}_{;\gamma}}^\tau (p_2)+\frac{13}{4}{\alpha^\prime}^{4}e_{\mu\nu;\rho\tau}(p_1)e^{\mu\nu;\rho\tau}(p_2)
\end{eqnarray}

\section{Regge behaviour in warped space}
In this section we revisited the arguments of $(\cite{polchinski})$ to investigate about the Regge limit for the analogous of (\ref{VS}) in the warped space background $ADS_5$. The metric is
\begin{equation}
ds^2 = \sqrt{\lambda}\frac{dy^2+\eta_{\mu\nu}dx^\mu dx^\nu}{y^2}.
\label{ads}
\end{equation}


Consider two vertex operators
$$V_1=e^{ip_1X}\psi_1(Y)\,\,\,\,\,\,V_2=e^{ip_2X}\psi_2(Y),
\label{states}$$
$X(z,\bar{z})$ and $Y(z,\bar{z})$ are the coordinates of the string and so the world-sheet fields. Following \cite{poly}, \cite{tseytlin} we assume that at the lowest order in the world-sheet perturbation theory, the world-sheet supersymmetry as well the Ramond-Ramond fluxes do not make any contribution. So in the following the dynamics will be given by the non-linear $\sigma$-model
\begin{equation}
S=\frac{\sqrt{\lambda}}{4\pi \alpha^\prime}\int d^2z \left[\frac{\partial X^\mu \bar{\partial} X^\nu \eta_{\mu\nu}}{Y^2}+\frac{\partial Y \bar{\partial}Y}{Y^2}\right].
\label{sigma}
\end{equation}

To perform the OPE we naively multiply the vertex operators, keeping only those terms that survive in the Regge limit:
\begin{equation}
V_1(z,\bar{z})V_2(0,0)=exp[i(p_1+p_2)X(0,0)+ip_1 (z\partial X(0,0)+\bar{z}\bar{\partial}X (0,0))]\psi_1(Y(0,0))\psi_2(Y(0,0))
\end{equation}
and with $k=p_1+p_2$ the first two terms giving a nonul integral $\int d^2z$ on the entire complex plane are
\begin{equation}
exp[ik X(0,0)][1-|z|^2 p_{1\mu}p_{1\nu}\partial X^\mu\partial X^\nu+......]\psi_1(Y(0,0))\psi_2(Y(0,0)),
\label{two}
\end{equation}
that we rewrite as 
\begin{equation}
exp[ik x][1-|z|^2 p_{1\mu}p_{1\nu}\partial X^\mu\partial X^\nu+......]\psi_1(y)\psi_2(y)
\end{equation}
meaning by the lower cases the zero modes of the world-sheet fields. In the Regge limit the second term in the sum (\ref{two}) is the dominant one, we may therefore take by following \cite{polchinski}
\begin{eqnarray}
&&-|z|^2 exp[ik X(0,0)]p_{1\mu}p_{1\nu}\partial X^\mu\partial X^\nu \psi_1(Y(0,0))\psi_2(Y(0,0)) \sim \label{dominant}\\
&& exp[ik X(0,0)+ip_1 (z\partial X(0,0)+\bar{z}\bar{\partial}X (0,0))]\psi_1(Y(0,0))\psi_2(Y(0,0)),\nonumber
\end{eqnarray}
and correct our Regge trajectory in $\frac{1}{\sqrt{\lambda}}$ by simply renormalizing at one loop the result (\ref{dominant}). This procedure just consists into compute the one loop anomalous dimension of the tensor term in (\ref{dominant}) inside the $\sigma$ model  $(\ref{sigma})$.
The one-loop OPE is given by
\begin{equation}
|z|^{-4+D_{\rm op}}:exp[ik X(0,0)][1-|z|^2 p_{1\mu}p_{1\nu}\partial X^\mu\partial X^\nu+......]\psi_1(Y(0,0))\psi_2(Y(0,0)):
\label{taylor}
\end{equation} 
the $-4$ in the Wilson coefficient exponent is because the vertex operators $V_1$ and $V_2$ are on-shell. $D_{\rm op}$ is the anomalous dimension operator for ($\ref{dominant}$). To determine such an operator acting on a given vertex operator $\Omega (z,\bar{z})$ we use the standard perturbative renormalization theory techniques. The change $\dot{\Omega}$ results by applying an external Liouville field $\sigma$ in the usual conformal rescaling of the world-sheet metric $\eta_{\alpha\beta}\longrightarrow \gamma_{\alpha\beta}=e^{2\sigma}\eta_{\alpha\beta}$. In the limit of small curvature of the $ADS_5$ space or, wich is the same, in the limit of strong coupling $\lambda$, related each others by \cite{malda}
\begin{equation}
R^2=\sqrt{\lambda} \alpha^\prime,
\end{equation}
the anomalous dimension for the tensor operator in the sum ($\ref{two}$) has been computed in \cite{CALLANGAN}. Adapting the result of that paper to our case we obtain 
\begin{equation}   
|z|^{-4+\alpha^\prime(D_1+\frac{k^2 y^2}{\sqrt{\lambda}})} \{ \psi_1(y)\psi_2(y)\}exp[ik X(0,0)+ip_1 (z\partial X(0,0)+\bar{z}\bar{\partial}X (0,0))].
\label{OPE}
\end{equation}
The operator $D_1$ is defined as 
\begin{equation}
\frac{\delta}{\delta\sigma}<F_{MN}(Z)\gamma^{\alpha\beta}\partial_\alpha Z^M \partial_\beta Z^N>_{Z_0}=\alpha'\left(D_1 +\frac{k^2 y^2}{\sqrt{\lambda}}\right) [F_{MN}(Z)]\gamma^{\alpha\beta}\partial_\alpha {Z_0}^M \partial_\beta {Z_0}^N
\end{equation}
where in our case $Z=(X,Y)$ and

$$F_{\mu\nu}= p_{1\mu}p_{1\nu}e^{i(p_1+p_2)\cdot X}\psi_1(Y(0,0))\psi_2(Y(0,0))$$
$$F_{\mu y}=0\,\,\,\,\,\,\,\,\,\, F_{yy}=0$$ 

The result of \cite{CALLANGAN} that we are going to apply is
 \begin{eqnarray}
&&\frac{\delta}{\delta\sigma}<F_{MN}(Z)\gamma^{\alpha\beta}\partial_\alpha Z^M \partial_\beta Z^N>_{Z_0}=\nonumber\\ 
&&\alpha^\prime \{-\nabla^2 F_{MN}-\nabla_M \nabla_N {F^L}_L +{{R_M}^{LS}}_N F_{LS} +2\nabla_M\nabla^L F_{LN}+ 2 \nabla_N \nabla^L F_{LM}\}\gamma^{\alpha\beta}\partial_\alpha {Z_0}^M \partial_\beta {Z_0}^N\nonumber\\
&& (\alpha^\prime)^2  \left\{\frac{1}{4} \nabla^2 {F^L}_L - \frac{1}{4}\nabla^L \nabla^S F_{LS} +\frac{1}{4} R^{LS}F_{LS}\right\}\,\,\, ^{(2)}R
\label{NPB}
\end{eqnarray}

here the Laplacian $\nabla^2$, the covariant derivatives $\nabla_M$, the Riemann tensor $R_{MNPQ}$ and the Ricci tensor $R_{MN}$ are computed for the $ADS_5$ space with the metric $(\ref{ads})$. $^{(2)}R$ is the worldsheet curvature of the Liouville rescaled metrics $\gamma_{\alpha\beta}=e^{2\sigma}\eta_{\alpha\beta}$. Remembering that $ADS_5$ is a maximally symmetric space its Riemann tensor is given by
\begin{equation}
R_{MNPQ}=\frac{\Lambda}{12}(G_{MP}G_{NQ}-G_{MQ}G_{NP})
\end{equation}
$\Lambda$ being a negative constant. Our $F_{MN}$ provides a vertex operator which is "multiplicatively renormalizable ", because the mixing with world-sheet curvature $^{(2)}R$ is easily seen  
$$R_{LS}F^{LS}\propto \frac{p_1^2}{\sqrt{\lambda}},$$
which is vanishing, because the quadrimensional masses can be neglected in the Regge limit
$$p_1^2 \rightarrow 0\,\,\,\,\,\,\, p_2^2 \rightarrow 0.$$
Since $(\ref{decomposition})$ the transfered momentum $k=p_1+p_2$ is orthogonal to $p_1$ and $p_2$ therefore 
$$(2\nabla_M\nabla^L F_{LN}+ 2 \nabla_N \nabla^L F_{LM})\longrightarrow 0$$
so the  OPE reduces to (\ref{OPE}) with $D_1$ given by
\begin{equation}
D_1=\frac{y^2 k^2 -y^2 {\partial^2}_y -2y\partial_y +2+\frac{\Lambda\sqrt{\lambda}}{12}}{\sqrt{\lambda}}.
\label{anomalous}
\end{equation}
By replacing (\ref{OPE}) and (\ref{anomalous}) in the matrix element 
\begin{equation}
<V_1(z,\bar{z})V_2(0,0)V_3(1)V_4(\infty)> 
\end{equation}
we get 
\begin{equation}
F(y)<e^{ik X (0,0)+ i{p_1}(z\partial X(0,0)+\bar{z}\bar{\partial}X (0,0)))}e^{ip_3 X(1)}e^{ip_4 X(\infty)}>
\label{semi}
\end{equation}
being
\begin{equation}
F(y)=|z|^{-4+\alpha^\prime (D_1+\frac{k^2 y^2}{\sqrt{\lambda}})} \{\psi_1(y)\psi_2(y)\}.
\end{equation}
The matrix element (\ref{semi}) is computed by making the semiclassical Wick contractions of the $X$ fields by using
\begin{equation}
-\frac{\alpha^\prime}{2}\frac{y^2}{\sqrt{\lambda}}\eta^{\mu\nu}ln|z_1-z_2|^2,
\end{equation}
and the result is 
\begin{equation}
F(y)e^{-\frac{\alpha^\prime p_1 p_3y ^2}{2\sqrt{\lambda}}(z+\bar{z})}.
\end{equation}
The amplitude is obtained as a coherent superposition of all the matrix elements $(\ref{semi})$ eavaluated at all the transverse position $y$. Moreover we have to perform the $z$ integration of our matrix element, meaning the calculation of the integral 
\begin{equation}
\int dy \sqrt{-G}\psi_3(y)\psi_4(y)\int d^2 z e^{-\frac{\alpha^\prime s y^2}{4\sqrt{\lambda}}(z+\bar{z})}|z|^{-4+\alpha^\prime(D_1 +\frac{k^2 y^2}{\sqrt{\lambda}})}\psi_1(y)\psi_2(y)
\end{equation}
proceeding as in the previous section we obtain
\begin{equation}
\int dy \sqrt{-G}\psi_3(y)\psi_4(y)\Pi\left[-2\alpha^\prime\left(D_1 +\frac{k^2 y^2}{\sqrt{\lambda}}\right)\right]\left(\frac{-i\alpha^\prime y^2 s}{4\sqrt{\lambda}}\right)^{2-\alpha^\prime\left(D_1 +\frac{k^2 y^2}{\sqrt{\lambda}}\right)}\psi_1(y)\psi_2(y)),
\label{result}
\end{equation}
meaning 
$$\Pi(x)=2\pi e^{-i\pi-i\pi\frac{x}{4}}\frac{\Gamma(-1-\frac{x}{4})}{\Gamma(2+\frac{x}{4})}.$$

\section{Conclusions and discussions}
In this paper we performed an analysis of the pomeron in closed string theory. The couplings of the pomeron vertex operator to some of the closed string states have been explicitly computed. The utility of such quantities consists into providing the asymptotic Regge behaviour of the scattering amplitude of a two by two process. Giving the coupling $(\ref{coupling})$, the two by two scattering amplitude $(\ref{VS})$ is reduced to the computation of the three point Green function
\begin{equation}
C_{12}<0|V_P(0)V_3(1)V_4(\infty)|0>.
\end{equation} 
Regge behaviour in a warped space like $ADS_5$ is still not fully understood and the result $(\ref{result})$ for our scattering amplitude is done in some semiclassical approximation. The qualitative meaning is not so different from the one of \cite{polchinski} neverthless the methods of derivation are not the same. In this paper the calculation of the anomalous dimension of the massless vertex operator as done by \cite{CALLANGAN} has been used assuming that the two derivative operator gives the dominant contribution in the Regge operator product expansion $(\ref{OPE})$. This is the same assumption as in \cite{polchinski}, neverthless we do not make use of the basis of the many derivatives operators 
\begin{equation}
V(j))=(\bar{\partial}X^+ \partial X^+)^{\frac{j}{2}}e^{ikX}\varphi_{+^j}.
\end{equation}
since for $j$ different than zero and two a complicated mixing is expected in the one loop world-sheet renormalization with a very complicated anomalous dimension operator. Of course one could think about a redifined basis of multiplicatively renormalized operators but the anomalous dimension operator is not known \cite{CALLANGAN}. Assuming by dimensional analysis that the anomalous dimension for the operator $V(j)$ was 
\begin{equation}
L_0 V(j)=(\partial{X^+}\bar{\partial}{X^+})^{\frac{j}{2}} \left[\frac{j}{2}-\frac{\alpha^\prime}{4}(\Delta_j+\delta_j)\right]e^{ikX}\varphi_{+^j}
\end{equation}
where $\Delta_j$ is the covariant Laplacian for the $j$-tensor $\varphi_{+^j}$, we have found $\delta_2 \neq 0$ due to the contribution in (\ref{NPB}) from the Reimann tensor. This is a small shift in the Regge intercept which can be neglected for an $ADS_5$ space with small curvature. 
Probably something more can be said by studying the problem in the global coordinates  \cite{tseytlin} where the $ADS_5$ sigma model $(\ref{sigma})$ appears as an $O(5,1)$ non linear sigma model.  

\section*{Acknowledgments}
We would like to acknowledge in advance the readers who kindly will send us their comments in order to improve our understanding about this subject.

\end{document}